\documentclass[
 reprint,
 twocolumn,
 superscriptaddress,
amsmath,amssymb,
aps,
]{revtex4-2}
\usepackage{amsthm}
\usepackage{amsfonts}
\usepackage{tikz}
\usepackage{ascmac}
\usepackage{braket}
\usepackage[colorlinks, linkcolor=blue, citecolor=blue, urlcolor=blue]{hyperref}
\newtheorem{theorem}{Theorem}

\newtheorem*{theorem*}{Theorem} 
\newtheorem{lemma}[theorem]{Lemma}

\newtheorem*{lemma*}{Lemma}
\newtheorem{definition}[theorem]{Definition}
\newtheorem*{definition*}{Definition}
\newtheorem*{conjecture*}{Conjecture}

\begin{document}

\title{Nonlinear transformation of complex amplitudes via quantum singular value transformation}

\author{Naixu Guo}
 \email{naixuguo@gmail.com}
 \affiliation{Graduate School of Engineering Science, Osaka \ University, 1-3 \  Machikaneyama, Toyonaka, Osaka \  560-8531, Japan}
\author{Kosuke Mitarai}
 \email{mitarai@qc.ee.es.osaka-u.ac.jp}
 \affiliation{Graduate School of Engineering Science, Osaka \ University, 1-3 \  Machikaneyama, Toyonaka, Osaka \  560-8531, Japan}
 \affiliation{
  Center for Quantum Information and Quantum Biology,
  Osaka\  University,\ Japan}
  \affiliation{JST PRESTO, Japan}
\author{Keisuke Fujii}
 \email{fujii@qc.ee.es.osaka-u.ac.jp}
 \affiliation{Graduate School of Engineering Science, Osaka \ University, 1-3 \  Machikaneyama, Toyonaka, Osaka \  560-8531, Japan}
 \affiliation{
  Center for Quantum Information and Quantum Biology,
  Osaka\  University,\ Japan}
  \affiliation{Center for Emergent Matter Science, RIKEN, Wako Saitama 351-0198, Japan}
\date{\today}

\begin{abstract}
Due to the linearity of quantum operations, it is not straightforward to implement nonlinear transformations on a quantum computer, making some practical tasks like a neural network hard to be achieved. 
In this work, we define a task called \textit{nonlinear transformation of complex amplitudes}
and provide an algorithm to achieve this task.
Specifically, we construct a block-encoding of complex amplitudes from a state preparation unitary.
This allows us 
to transform the complex amplitudes by using quantum singular value transformation. 
We evaluate the required overhead in terms of input dimension and precision,
which reveals that the algorithm depends on the roughly square root of input dimension and achieves an exponential speedup on precision compared with previous work.
We also discuss its possible applications to quantum machine learning, where complex amplitudes encoding classical or quantum data are processed by the proposed method.
This paper provides a promising way to introduce the highly complex nonlinearity of the quantum states, which is essentially missing in quantum mechanics.
\end{abstract}

\maketitle

\section{Introduction}

Quantum mechanics is fundamentally linear.
Quantum operations are all unitaries in matrix representation, and the measurement is the only way in which we can produce the nonlinearity. 
If deterministic nonlinear operations are allowed, quantum computers can solve NP-hard problems in polynomial time \cite{PhysRevLett.81.3992}.
This result implies the difficulty of nonlinear quantum operations.
However, nonlinearity plays an essential role in applications such as nonlinear differential equations and machine learning.
It is important to consider how to implement a nonlinear transformation on a quantum computer, even if it can only be applied probabilistically.

The key idea for applying probabilistic nonlinear operations is to use a post-selection, i.e., the desired transformation is applied on target subsystems if ancilla qubits are measured in a certain state.
The Harrow-Hassidim-Lloyd (HHL) algorithm \cite{PhysRevLett.103.150502}, for example, used this idea to provide an exponential speedup for solving linear systems of equations compared with classical computers.

In recent years, methods for performing nonlinear operations have been developed rapidly.
Quantum signal processing \cite{PhysRevLett.118.010501} is a technique to perform desired nonlinear functions on eigenvalues of a unitary operator. 
Quantum singular value transformation (QSVT) \cite{10.1145/3313276.3316366} is its generalization to perform nonlinear functions on singular values of a matrix encoded in a unitary.
Existing quantum algorithms can be reconstructed in terms of QSVT \cite{10.1145/3313276.3316366, martyn2021grand}, which unifies the disparate problems like Hamiltonian simulation \cite{10.5555/2481569.2481570, PhysRevLett.114.090502, PhysRevLett.118.010501, Low_2019}, linear equation solving \cite{PhysRevLett.103.150502, Childs_2017} and amplitude amplification techniques \cite{Berry_2015, PhysRevLett.113.210501, low2017hamiltonian}, implying the importance of developing tools to introduce nonlinearity on quantum computers.

Here, we consider how to perform nonlinear transformations on the state amplitudes directly, rather than the nonlinear transformation of matrices described in the previous work \cite{10.1145/3313276.3316366}. 
We call this task \textit{nonlinear transformation of complex amplitudes} (NTCA), which is defined as follows.
Assume we have access to a state preparation oracle 
\begin{align}
    U:\ket{0}\rightarrow\sum_{k=1}^N c_k\ket{k},    
\end{align}
which encodes a complex vector $\Vec{c}=\{c_1,\cdots,c_N\}$.
We also assume having access to its adjoint and their controlled versions.
For given nonlinear functions $P,Q:\mathbb{R}\rightarrow\mathbb{C}$, NTCA is a task to prepare a quantum circuit such that its output state is 
\begin{align}
    \frac{1}{\sqrt{c}}\sum_{k=1}^N \left(P(x_k)+Q(y_k)\right)\ket{k},
\end{align}
where $x_k+iy_k=c_k$ and $c$ is a normalization factor.

In this work, we provide an algorithm for NTCA based on QSVT.
To achieve this, we develop a method called \textit{block-encoding of amplitudes}, where a unitary operator constructed from the state preparation oracle encodes a Hermitian matrix, whose singular values are the real or imaginary parts of amplitudes, into its subspace.
We show that if functions $P$ and $Q$ can be approximated by $d$-degree polynomials $P'$ and $Q'$, NTCA can be implemented using controlled-$U$ and controlled-$U^\dagger$ $$\mathcal{O}\left(d\sqrt{N/\sum_{k=1}^N |P'(x_k)+Q'(y_k)|^2}\right)$$ times in expectation.
We also provide a discussion about how to exploit this algorithm for quantum machine learning.
More specifically, we consider how to perform a neural network on the quantum computer and discuss its possible applications.
This work opens up a way to implement nonlinear operations on a quantum computer and may contribute to the field of quantum machine learning.

This paper is organized as follows.
We define the task NTCA and summarize existing important subroutines in Section \ref{section:2}.
In Section \ref{section:3}, we introduce the \textit{block-encoding of amplitudes}.
Combining this method with QSVT, we construct an algorithm to perform NTCA in Section \ref{section:4}.
In Section \ref{section:5}, we provide the application of this algorithm and show that NTCA provides us a way to perform the neural network task directly on a quantum computer.

\section{Preliminary \label{section:2}}
In this section, we first define a task called NTCA and summarize some subroutines from existing works. Throughout this paper, $N$ denotes the dimension of data, $n=\lceil\log_2(N)\rceil$, and $\mathcal{I}_{s}$ denotes the identity operator for $s$ qubits.
A quantum state $\ket{\psi}$ represented in the form $\ket{\psi}=c_a\ket{a}+\cdots$ means that we focus on the term $c_a\ket{a}$ and the residual term is orthogonal to the first term.

First, we formally define NTCA as follows:

\begin{definition}[\label{NTCA}Nonlinear transformation of complex amplitudes (NTCA)]
    Assume we have access to a state preparation oracle $U$ such that $U\ket{0}=\sum_{k=1}^N c_k\ket{k}$, where $c_k=x_k+i y_k$ and $x_k,y_k$ are real numbers.
    We also assume having access to its adjoint and their controlled versions.
    For nonlinear functions $P,Q:\mathbb{R}\rightarrow\mathbb{C}$ bounded on the interval $[-1,1]$,
    output a quantum state $\frac{1}{\sqrt{c'}}\sum_{k=1}^N b_k\ket{k}$ such that it is an $\varepsilon$-approximation to the $\frac{1}{\sqrt{c}}\sum_{k=1}^N \left(P(x_k)+Q(y_k)\right)\ket{k}$.
    Here, $c,c'$ are the normalization coefficients, and $\varepsilon$-approximation means that, for all $k$ and $x_k,y_k\in [-1,1]$,
    \begin{align}
         |b_k-P(x_k)-Q(y_k)|\leq \frac{\varepsilon}{N}.
    \end{align}

\end{definition}

To achieve NTCA, we utilize QSVT, which is a technique to apply nonlinear functions on singular values of a matrix.
It exploits the so-called block-encoding of a target matrix defined as follows.
\begin{definition}[\label{blkencod}Block-encoding \cite{Low_2019, chakraborty_et_al:LIPIcs:2019:10609}]
    Suppose that A is an n-qubit operator, $\alpha, \varepsilon\in\mathbb{R}_{+}$ and $a\in \mathbb{N}$. Then we say that the $(a+n)$-qubit unitary $U$ is an $(\alpha,a,\varepsilon)$-block-encoding of A if
    \begin{align}
        \|A-\alpha(\bra{0}^{\otimes a}&\otimes \mathcal{I}_n)U(\ket{0}^{\otimes a}\otimes \mathcal{I}_n)\|\leq \varepsilon.
    \end{align}
\end{definition}
\noindent Using the block-encoding, QSVT can implement $P(A)$ of a Hermitian matrix $A$ for any polynomial $P$. More formally, we can achieve the following.

\begin{lemma}[Polynomial eigenvalue transformation of arbitrary parity {\cite[Theorem 56]{10.1145/3313276.3316366}\label{QSVT}}]
Suppose that $U$ is an $(\alpha, a, \varepsilon)$-block-encoding of a Hermitian matrix $A$.
If $\delta \geq 0$ and $P:\mathbb{R}\rightarrow \mathbb{C}$ is a d-degree polynomial satisfying that
\begin{align}
    \mathrm{for\ all}\ x \in[-1,1]:\left|P(x)\right| \leq \frac{1}{4},
\end{align}
then there is a quantum circuit $\tilde{U}$, which is an $(1, a+3, 4d \sqrt{\varepsilon / \alpha}+N\delta)$-block-encoding of $P(A / \alpha)$, and consists of d applications of $U$ and $U^{\dagger}$ gates, a single application of controlled-U and $\mathcal{O}(ad)$ other one- and two-qubit gates. Moreover, we can compute a description of such a circuit with a classical computer in time $\mathcal{O}(\operatorname{poly}(d, \log (1 / \delta)))$.
\end{lemma}
We note that this theorem is not exactly Theorem 56 of Ref. \cite{10.1145/3313276.3316366} but a slightly generalized version which is mentioned in Ref. \cite{10.1145/3313276.3316366}.
The original theorem is a similar claim for real polynomial $P_\Re:\mathbb{R}\to\mathbb{R}$ requiring two ancilla qubits instead of three in the above theorem.

\section{Result}
\subsection{Block-encoding of amplitudes \label{section:3}}
In this subsection, we describe a method that we call \textit{block-encoding of amplitudes}.
Using the state preparation oracle $U$ as Definition \ref{NTCA}, 
we construct $(1,1,0)$-block-encodings $\tilde{G}$ and $\tilde{G'}$ of $(2n+1)$-qubit Hermitian matrices, whose eigenvalues include respectively the real and imaginary part of amplitudes, i.e., 
\begin{align}\label{BCFA1}
    (\bra{0}\otimes \mathcal{I}_{2n+1})\tilde{G}(\ket{0}\otimes\mathcal{I}_{2n+1})=\sum_{k=1}^N x_k |\varphi_k\rangle\langle\varphi_k |+ \cdots,\\ \label{BCFA2}
    (\bra{0}\otimes \mathcal{I}_{2n+1})\tilde{G'}(\ket{0}\otimes\mathcal{I}_{2n+1})=\sum_{k=1}^N y_k |\varphi'_k\rangle\langle\varphi'_k |+\cdots,
\end{align}
where $\ket{\varphi_k}$ and $\ket{\varphi'_k}$ are mutually orthogonal $(2n+1)$-qubit states. 
Concrete forms of $\ket{\varphi_k}$ and $\ket{\varphi'_k}$ will be described later.
The result is summarized as the following theorem.

\begin{theorem}[Block-encoding of real part amplitudes\label{BCA}]
    Given a state preparation oracle $U$ as Definition \ref{NTCA},
    we can construct a unitary $\tilde{G}$ and $\tilde{G}'$ which satisfy Eq. (\ref{BCFA1}) and (\ref{BCFA2}) respectively by using controlled-$U$, controlled-$U^\dagger$ four times, and $\mathcal{O}(n)$ one- and two-qubit gates.
\end{theorem}

\begin{figure}
    \centering
    \includegraphics[width=\linewidth]{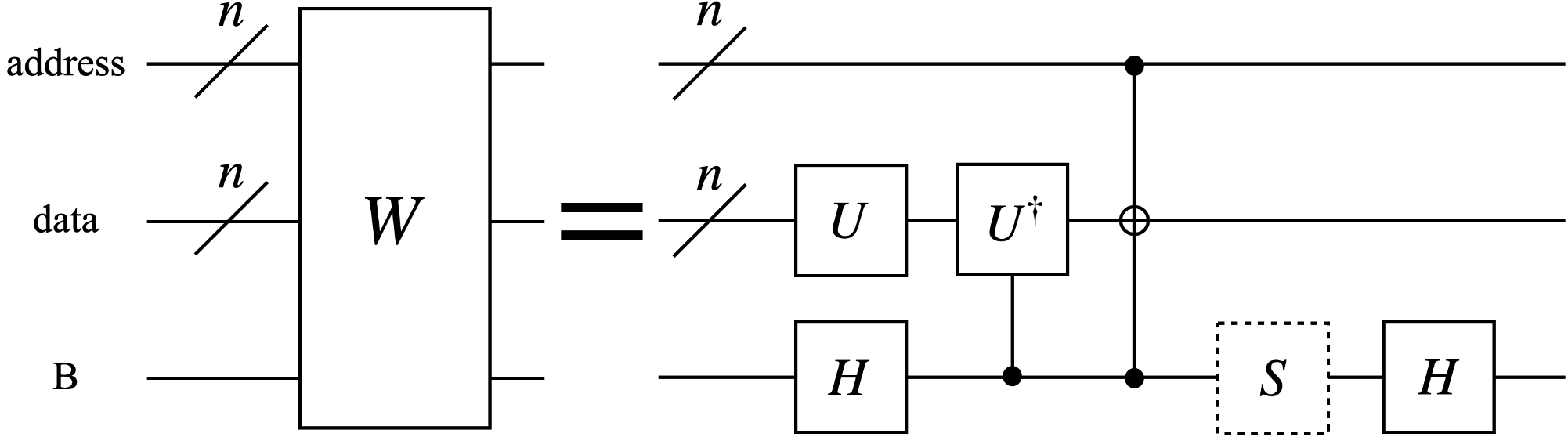}
    \caption{Definition of operation $W$ in Theorem \ref{BCA}}
    \label{fig:11}
\end{figure}
\begin{figure}
    \centering
    \includegraphics[width=\linewidth]{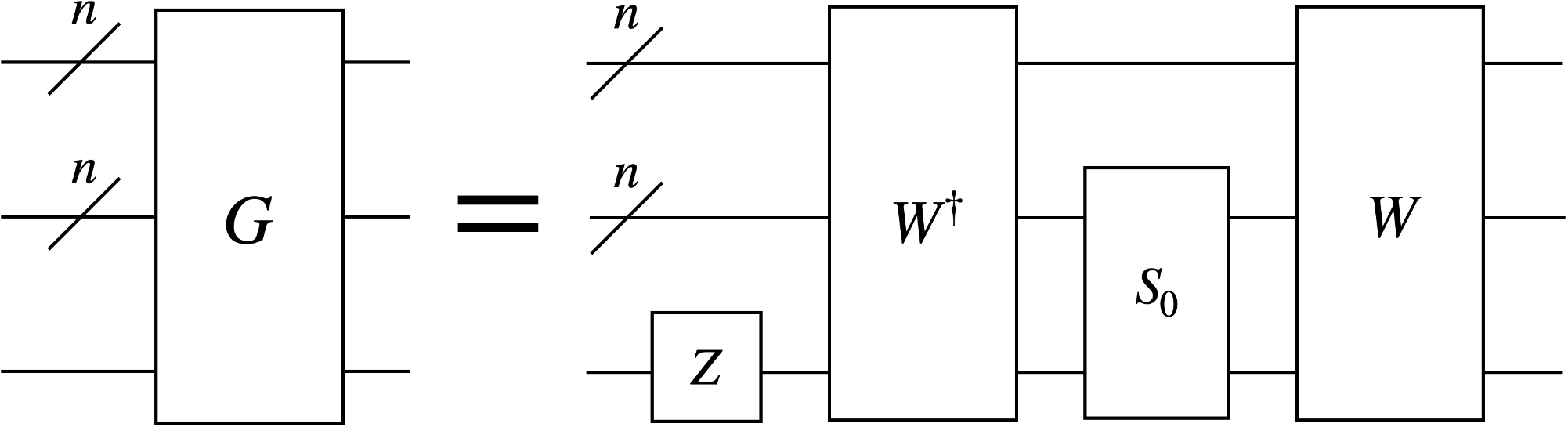}
    \caption{Definition of operation $G$ in Theorem \ref{BCA}}
    \label{fig:12}
\end{figure}
\begin{figure}
    \centering
    \includegraphics[width=\linewidth]{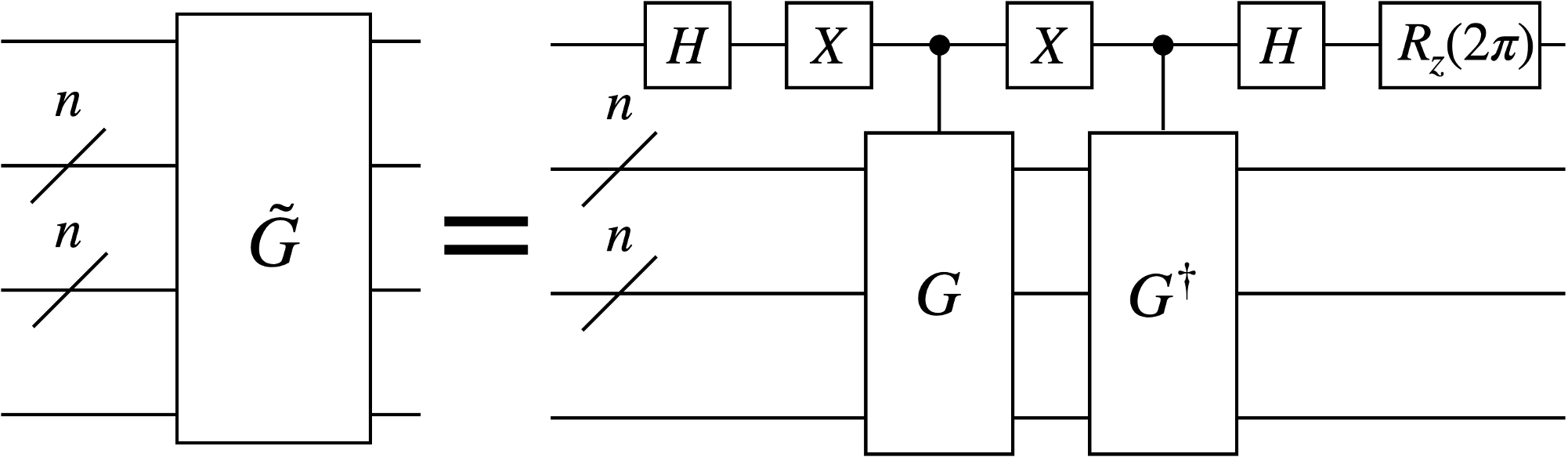}
    \caption{Definition of operation $\tilde{G}$ in Theorem \ref{BCA}}
    \label{fig:13}
\end{figure}

\begin{proof}
The construction is partly based on the previous work \cite{Mitarai_2019}.
First, we construct a unitary whose eigenvalues encode the information of $x_k$.
Let $W$ be a sequence of unitary operations as shown in Fig. \ref{fig:11}.
Note that,
\begin{align}
    W\ket{k}_{\mathrm{ad}}\ket{0}_{\mathrm{da,B}} &= \frac{\ket{k}}{2}_{\mathrm{ad}}\biggl(\bigl(\sum_{j=1}^N c_j\ket{j}+\ket{k}\bigl)_{\mathrm{da}}\ket{0}_{\mathrm{B}}\notag\\
        &+\bigl(\sum_{j=1}^Nc_j\ket{j}-\ket{k}\bigl)_{\mathrm{da}}\ket{1}_{\mathrm{B}}\biggl). \label{eq:op_W}
\end{align}
See Appendix \ref{app.op.W} for details.
Let $S_0:=\mathcal{I}_{n+1}-2|0\rangle\langle 0|_{\mathrm{da,B}}$ be a conditional phase-shift gate on the data qubits and ancilla qubit B, and $Z_{\mathrm{B}}$ be the $Z$ gate acting on the ancilla qubit B.
With these operations, we define 
\begin{align}\label{eq:def_G}
    G:=WS_0 W^\dagger Z_{\mathrm{B}},
\end{align}
which is shown in Fig. \ref{fig:12}.
For $k\in\{1,\dots,N\}$, we define the following normalized states
\begin{align}
    \ket{\Psi_{k0}}_{\mathrm{da,B}}:=\frac{1}{2\alpha_k}\biggl(&\sum_j c_j\ket{j}+\ket{k}\biggl)_{\mathrm{da}}\ket{0}_{\mathrm{B}}\label{eigenstat0}\\
    \ket{\Psi_{k1}}_{\mathrm{da,B}}:=\frac{1}{2\beta_k}\biggl(&\sum_j c_j\ket{j}-\ket{k}\biggl)_{\mathrm{da}}\ket{1}_{\mathrm{B}}\label{eigenstat1},
\end{align}
where $\alpha_k=\sqrt{\frac{1}{2}(1+x_k)}$, $\beta_k=\sqrt{\frac{1}{2}(1-x_k)}$.
As shown in Appendix \ref{app.op.G}, eigenvectors of $G$ are
\begin{align}\label{eq:eigenvector_G}
    \ket{k}_{\mathrm{ad}}\ket{\Psi_{k\pm}}_{\mathrm{da,B}}:=\frac{1}{\sqrt{2}}\ket{k}_{\mathrm{ad}}\left(\ket{\Psi_{k0}}\pm\ket{\Psi_{k1}}\right)_{\mathrm{da,B}},
\end{align}
with the corresponding eigenvalues $-x_k \pm i\sqrt{1-x_k^2}$.
We neglect other eigenstates of $G$ since we are only interested in the subspace which encodes amplitudes.
However, since our target is to achieve an operator as Eq. (\ref{BCFA1}), we need to remove the imaginary part of eigenvalues.

We achieve this by the circuit in Fig. \ref{fig:13} where we use controlled versions of $G$ and $G^\dagger$.
The idea is that the imaginary part of eigenvalues can be canceled by implementing $\frac{1}{2}(G+G^\dagger)$.
Define $\tilde{G}$ as shown in Fig. \ref{fig:13}.
We then have
\begin{align}
(\bra{0}\otimes\mathcal{I}_{2n+1})\ \tilde{G} \ (\ket{0}\otimes \mathcal{I}_{2n+1})=-\frac{1}{2}(G+G^\dagger).
\end{align}
Therefore, $\tilde{G}$ is a $(1,1,0)$-block-encoding of Hermitian matrix $-\frac{1}{2}(G+G^\dagger)$.
Eigenvalues of $-\frac{1}{2}(G+G^\dagger)$ are
$\{x_k\}_{k=1}^N$.
To construct $\tilde{G}$, we need to use controlled-$U$, controlled-$U^\dagger$ four times, and $\mathcal{O}(n)$ one- and two-qubit gates. 
Details of gate count are provided in Appendix \ref{app.num.gates}. 

$\tilde{G}'$ can be constructed by the same procedure except that we add an $S$ gate on the ancilla qubit $B$ in $W$ as in Fig. \ref{fig:11}.
\end{proof}

\subsection{Nonlinear transformation of complex amplitudes \label{section:4}}

In this subsection, we provide an algorithm to implement NTCA.
The idea is to apply Lemma \ref{QSVT} to the block-encodings obtained in Theorem \ref{BCA} to implement desired functions.
We assume the existence of polynomial approximations of functions $P$ and $Q$ to utilize Theorem \ref{QSVT}. 
Note that in principle, any continuous real-valued function defined on the interval $[-1,1]$ can be approximated by a polynomial function with arbitrary precision by the Weierstrass approximation theorem~\cite{PINKUS20001}.

Our quantum algorithm for NTCA is the following.

\begin{theorem}[Nonlinear transformation of complex amplitudes]
    \label{NTCAalg}
    If functions $P$ and $Q$ in Definition \ref{NTCA} can be approximated by $d_p$- and $d_q$-degree polynomials $P'$ and $Q'$ such that $|P'(x)-P(x)|, |Q'(x)- Q(x)|\leq \varepsilon/(4N)$ for all $x\in[-1,1]$, the following holds.
    Let $d:=\max \{d_p,d_q\}$ and $\gamma:=\underset{x\in [-1,1]}{\max}\{|P(x)|,|Q(x)|\}$, then
    NTCA in Definition \ref{NTCA} can be achieved by using controlled-$U$, controlled-$U^\dagger$ $$\mathcal{O}\left(d\gamma\sqrt{N/\sum_{k=1}^N |P'(x_k)+Q'(y_k)|^2}\right)$$ times, and $$\mathcal{O}\left(nd\gamma\sqrt{N/\sum_{k=1}^N |P'(x_k)+Q'(y_k)|^2}\right)$$ one- and two-qubit gates in expectation. 
\end{theorem}
\noindent \textit{Proof overview.} We implement $P'(x)$ and $Q'(y)$ separately by applying Lemma \ref{QSVT} to the block-encodings obtained in Theorem \ref{BCA}.
This process provides us two unitaries $\mathcal{P}$ and $\mathcal{Q}$ which block-encodes matrices whose eigenvalues are $P'(x_k)/(4\gamma)$ and $Q'(y_k)/(4\gamma)$, respectively.
The coefficient of $4\gamma$ is necessary to make the functions implementable with Lemma \ref{QSVT}.
Noting that the eigenstates of the above two unitaries can be prepared by applying $W$ operation in Fig. \ref{fig:12} to a computational basis, we then apply $\mathcal{P}$ and $\mathcal{Q}$ to uniform superpositions of $N$ eigenstates corresponding to eigenvalues $P'(x_k)/(4\gamma)$ and $Q'(y_k)/(4\gamma)$.
The concrete algorithm is provided below.

\begin{proof}
First, note that $P'(x)/(4\gamma)$ satisfies $|P'(x)/(4\gamma)| \leq 1/4$ therefore compatible with Lemma \ref{QSVT}.
By using $\tilde{G}$ in Theorem \ref{BCA} as an input to Lemma $\ref{QSVT}$ and setting $\delta=\varepsilon/(16\gamma N)$ in Lemma \ref{QSVT}, we can construct the unitary $\mathcal{P}$ as a $(1, 4,\varepsilon/(16\gamma))$-block-encoding of matrix $P'\left(-\frac{1}{2}(G+G^\dagger)\right)/(4\gamma)$ where $G$ is the unitary defined by Eq.~(\ref{eq:def_G}).
Details of error analysis are provided in Appendix \ref{app.err.any}.
With a similar procedure, we can construct $\mathcal{Q}$ as a $(1, 4,\varepsilon/(16\gamma))$-block-encoding of matrix $Q'(-\frac{1}{2}(G'+G^{'\dagger}))/(4\gamma)$.

We can extract $P'(x_k)$ and $Q'(y_k)$ to the amplitudes by injecting eigenstates of $P'\left(-\frac{1}{2}(G+G^\dagger)\right)$ and $Q'(-\frac{1}{2}(G'+G^{'\dagger}))$ to $\mathcal{P}$ and $\mathcal{Q}$ respectively.
We claim that $W\ket{k}_{\mathrm{ad}}\ket{0}_{\mathrm{da,B}}$ and $W'\ket{k}_{\mathrm{ad}}\ket{0}_{\mathrm{da,B}}$ is an eigenstate of $\mathcal{P}$ and $\mathcal{Q}$, respectively.
This is because the two states $\ket{k}_{\mathrm{ad}}\ket{\Psi_{k\pm}}_{\mathrm{da,B}}$ in Eq. (\ref{eq:eigenvector_G}) are the eigenstates of $-\frac{1}{2}(G+G^\dagger)$ with the (degenerate) eigenvalue $x_k$.
Therefore, any linear combinations of $\ket{\Psi_{k\pm}}_{\mathrm{da,B}}$, or equivalently those of $\ket{\Psi_{k0}}_{\mathrm{da,B}}$ and $\ket{\Psi_{k1}}_{\mathrm{da,B}}$, are eigenvectors with eigenvalues $x_k$.
From Eq. (\ref{eq:op_W}), we see that $W\ket{k}_{\mathrm{ad}}\ket{0}_{\mathrm{da,B}}$ is a linear combination of $\ket{\Psi_{k0}}_{\mathrm{da,B}}$ and $\ket{\Psi_{k1}}_{\mathrm{da,B}}$ and hence an eigenvector with the eigenvalue $x_k$.
Therefore, we have,
\begin{align}
\begin{split}
    &\mathcal{P}\left[\ket{0}^{\otimes 4}\left(W\ket{k}_{\mathrm{ad}}\ket{0}_{\mathrm{da,B}}\right)\right] \\
    \quad&= \frac{P'(x_k)}{4\gamma}\left[\ket{0}^{\otimes 4}\left(W\ket{k}_{\mathrm{ad}}\ket{0}_{\mathrm{da,B}}\right)\right]+\cdots, \label{eq:mathcalP-action}
\end{split}\\
    \begin{split}
        &\mathcal{Q}\left[\ket{0}^{\otimes 4}\left(W'\ket{k}_{\mathrm{ad}}\ket{0}_{\mathrm{da,B}}\right)\right] \\
    \quad &= \frac{Q'(y_k)}{4\gamma}\left[\ket{0}^{\otimes 4}\left(W\ket{k}_{\mathrm{ad}}\ket{0}_{\mathrm{da,B}}\right)\right]+\cdots. \label{eq:mathcalQ-action}
    \end{split}
\end{align}
By injecting the uniform superposition of the eigenstates, i.e., $\frac{1}{\sqrt{N}}\sum_{k=1}^N \ket{0}^{\otimes 4}\left(W\ket{k}_{\mathrm{ad}}\ket{0}_{\mathrm{da,B}}\right)$ to $\mathcal{P}$, we can get 
\begin{align}
    \frac{1}{4\gamma\sqrt{N}}\sum_{k=1}^N P'(x_k)  \ket{0}^{\otimes 4}\left(W\ket{k}_{\mathrm{ad}}\ket{0}_{\mathrm{da,B}}\right),
\end{align}
which has transformed amplitudes.
We can do likewise to extract $Q'(y_k)$.

To achieve NTCA of Definition \ref{NTCA}, we have to take a linear combination of $P'(x_k)$ and $Q'(y_k)$.
This can be performed by the circuit in Fig. \ref{fig:3}. 
The state before the application of controlled-$\mathcal{P}$ gate in Fig. \ref{fig:3} is,
\begin{align}
\begin{split}\label{eq:NTCA_input}
    \frac{1}{\sqrt{2N}}\sum_{k=1}^{N}&\left[\ket{0}\ket{0}^{\otimes 4}\left(W\ket{k}_{\mathrm{ad}}\ket{0}_{\mathrm{da,B}}\right)\right.\\
    &\left.+\ket{1}\ket{0}^{\otimes 4}\left(W'\ket{k}_{\mathrm{ad}}\ket{0}_{\mathrm{da,B}}\right)\right].
\end{split}
\end{align}
Applying controlled-$\mathcal{P}$ and controlled-$\mathcal{Q}$ to this state leads to the following state:
\begin{align}
\begin{split}
    \frac{1}{4\gamma\sqrt{2N}}\sum_{k=1}^{N}&\left[P'(x_k)\ket{0}\ket{0}^{\otimes 4}\left(W\ket{k}_{\mathrm{ad}}\ket{0}_{\mathrm{da,B}}\right)\right.\\
    &\left.+Q'(y_k)\ket{1}\ket{0}^{\otimes 4}\left(W'\ket{k}_{\mathrm{ad}}\ket{0}_{\mathrm{da,B}}\right)\right] + \cdots.
\end{split}
\end{align}
Then, we apply controlled-$W^\dagger$, controlled-$W'^\dagger$ to un-compute the data and B qubits.
Finally, applying the Hadamard gate to the first qubit as in Fig. \ref{fig:3} provides us the desired state,
\begin{align}\label{statwithgarb}
    \frac{1}{8\gamma\sqrt{N}}\sum_{k=1}^{N}\left( P'(x_k)+Q'(y_k)\right)\ket{0}^{\otimes 5}\ket{k}_{\mathrm{ad}}\ket{0}_{\mathrm{da, B}}
    +\cdots,
\end{align}
which contains the target state neglecting the data and $B$ qubits.
It is obvious that we can implement NTCA if we measure the first five qubits and the outcome is $\ket{0}^{\otimes 5}$.
The probability of this event is $\frac{1}{64\gamma^2 N}\sum_k |P'(x_k)^2+Q'(y_k)^2|$.

\begin{figure}
    \centering
    \includegraphics[width=\linewidth]{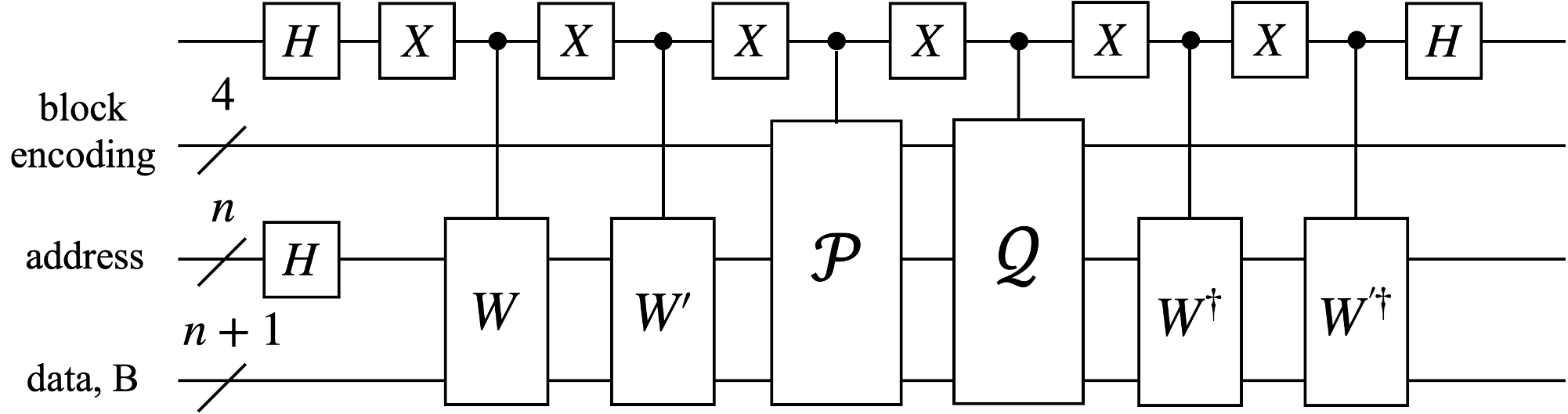}
    \caption{Quantum circuit for the nonlinear transformation of part of complex amplitudes}
    \label{fig:3}
\end{figure}

We can use amplitude amplification \cite{2000quant.ph..5055B} to boost this probability quadratically.
It allows us to achieve the NTCA with 
$$\mathcal{O}\left(\gamma\sqrt{\frac{N} {\sum_{k=1}^N |P'(x_k)+Q'(y_k)|^2}}\right)$$
use of the circuit in Fig. \ref{fig:3} in expectation.
Each of controlled-$\mathcal{P}$ and controlled-$Q$ operations is constructed with $\mathcal{O}(d)$ controlled-$U$ and controlled-$U^\dagger$ operations and $\mathcal{O}(nd)$ elementary gates.
Therefore, in total, we use controlled-$U$ and controlled-$U^\dagger$ 
$$\mathcal{O}\left(d\gamma\sqrt{\frac{N} {\sum_{k=1}^N |P'(x_k)+Q'(y_k)|^2}}\right)$$
times, and 
$$\mathcal{O}\left(nd\gamma\sqrt{\frac{N} {\sum_{k=1}^N |P'(x_k)+Q'(y_k)|^2}}\right)$$ 
one- and two-qubit gates in expectation to achieve NTCA.

\end{proof}

Notice that there is a natural generalization of Theorem \ref{NTCAalg}.
For $N_1\leq N$, we can output the state
\begin{align}\label{eq:partial_NTCA}
    \frac{1}{8\gamma\sqrt{N_1}}\sum_{k=1}^{N_1}(P'(x_k)+Q'(y_k))\ket{0}^{\otimes 5}\ket{k}+\cdots,
\end{align}
by slightly modifying the circuit in Fig. \ref{fig:3} so that the state just before the controlled-$\mathcal{P}$ operation is the following state, 
\begin{align}
    \begin{split}
    \frac{1}{\sqrt{2N_1}}\sum_{k=1}^{N_1}&\left[\ket{0}\ket{0}^{\otimes 4}\left(W\ket{k}_{\mathrm{ad}}\ket{0}_{\mathrm{da,B}}\right)\right.\\
    &\left.+\ket{1}\ket{0}^{\otimes 4}\left(W'\ket{k}_{\mathrm{ad}}\ket{0}_{\mathrm{da,B}}\right)\right],
    \end{split}
\end{align}
instead of the uniform superposition of $N$ eigenstates prepared in Eq. (\ref{eq:NTCA_input}).
If we are only interested in performing nonlinear transformations to a part of the amplitudes, we can benefit from this generalization.

Another remark is that, if we consider the case with real function $P:\mathbb{R}\to\mathbb{R}$ acting on the real part of amplitudes, we can reduce the total number of ancilla qubits from five to three. 
This is because we can construct $\mathcal{P}$ with only two ancilla qubits in this case as mentioned earlier in Sec. \ref{section:2}, and the sum between functions $P$ and $Q$ becomes unnecessary.
Therefore, with a continuous real-valued function $P$ and its polynomial approximation $P'$, for $N_1\leq N$ we can prepare the following state,
\begin{align}
    \frac{1}{2\gamma\sqrt{N_1}}\sum_{k=1}^{N_1} P'(x_k)\ket{0}^{\otimes 3}\ket{k}_{\mathrm{ad}}\ket{0}^{\otimes n+1}_{\mathrm{da,B}}+\cdots, \label{eq:real-ntca}
\end{align}
only requiring three ancilla qubits.

\subsection{Comparison with previous work}
Our work is inspired by the previous work \cite{Mitarai_2019}, based on which we focus on how to apply nonlinear functions on amplitudes of a quantum state on the quantum computer.
Ref. \cite{Mitarai_2019} also considered performing nonlinear transformation of amplitudes.
Let us compare the present work with Ref. \cite{Mitarai_2019} in the case where we wish to implement only $P:\mathbb{R}\to\mathbb{R}$ acting on real parts of amplitudes.
Their algorithm works by first performing quantum phase estimation \cite{Cleve_1998} of the unitary $G$ defined in Eq. (\ref{eq:def_G}) to encode $x_k$ (and $y_k$) into bitstrings as $\sum_{k=1}^N \ket{k}\ket{\bar{x}_k}$, where $\bar{x}_k$ is a binary representation of $x_k$.
Then, using an ancilla qubit and arithmetical operations, it prepares a state in the form of $$\sum_{k=1}^N \ket{k}\ket{\bar{x}_k}\left(P(x_k)\ket{0}+\sqrt{1-P(x_k)^2}\ket{1}\right),$$ from which we can obtain the desired state if we measure the ancilla qubit in $\ket{0}$ state.
If this algorithm is applied to the same setting as Definition \ref{NTCA}, it would require $\mathcal{O}(N/\varepsilon)$ uses of the controlled-$G$ and hence the same amount of controlled-$U$ in the phase estimation step.
In total, it would require $$\tilde{\mathcal{O}}\left(N/\varepsilon\sqrt{N/\sum_{k=1}^N |P(x_k)|^2}\right)$$ uses of controlled-$U$.
In contrast, the algorithm presented in Theorem \ref{NTCAalg} uses $$\mathcal{O}\left(d\gamma\sqrt{N/\sum_{k=1}^N |P'(x_k)+Q'(y_k)|^2}\right)$$ of controlled-$U$, where $d$ implicitly depends on the precision requirement $N/\varepsilon$.
In many cases, $d$ can be taken as $d=\mathcal{O}(\log(N/\varepsilon))$ \cite{10.1145/3313276.3316366}, and thus we achieve exponential speedup in terms of $N$ and $\varepsilon$.
This improvement is important when we consider the running time of possible applications like a quantum neural network, as described in the next section.
We also need fewer ancilla qubits for NTCA.
While our quantum algorithm needs $\mathcal{O}(n)$ ancilla qubits, Ref. \cite{Mitarai_2019} needs $\mathcal{O}(n+\log(N/\varepsilon))$.

\section{Application to quantum machine learning\label{section:5}}

Let us finally discuss possible applications of NTCA for machine learning tasks.
A straightforward application of the NTCA is to implement the neural network by using the complex amplitudes as its nodes as follows.
For a (classical) data vector $\Vec{x}=\{x_1,x_2,\cdots,x_N\} \in \mathbb{R}^N$, let us assume that we have a data-encoding unitary $U_{\Vec{x}}$ such that,
\begin{align}\label{uni.data}
    U_{\Vec{x}}:\ket{0}\rightarrow \sum_{k=1}^N x_k\ket{k}.
\end{align}
This is the so-called amplitude encoding of classical data \cite{Schuld2018book}.
$U_{\Vec{x}}$ can be constructed via the use of quantum random access memory (QRAM) \cite{PhysRevLett.100.160501} with the data structure mentioned in Ref. \cite{kerenidis2016quantum}.
Using $U_{\Vec{x}}$ and an orthogonal matrix $O=\sum_{j,k}o_{jk}|j\rangle\langle k|$, we can prepare,
\begin{align}
    \sum_{k=1}^N P\left(\sum_{j=1}^N o_{kj}x_j\right)\ket{k},
\end{align}
whose amplitudes correspond to the outputs of a single-layer neural network with an activation function $P$ and an orthogonal weight matrix, by using $OU_{\mathrm{data}}$ as $U$ in Theorem \ref{NTCAalg}.
Note that the normalization factor is omitted to simplify the notation.
We will do so henceforth if the normalization factor is not needed explicitly for discussions.
We also analyze the concrete implementation of a famous activation function  $P(x)=\tanh(x)$ in Appendix~\ref{app.appro.tanh}.

We can straightforwardly generalize them for a complex input vector $\ket{\psi}=\sum_{k=1}^N\psi_k\ket{k}$.
Let $U_\psi\ket{0}:=\ket{\psi}$.
Our NTCA algorithm can prepare a state in the form of,
\begin{align}\label{eq:quantumdata_nn}
    \sum_{k=1}^N F\left(\sum_{j=1}^Nv_{kj}\psi_j\right)\ket{k}
\end{align}
$F:\mathbb{C}\to\mathbb{C}$ is an activation function in the form of $F(\psi)=P(\mathrm{Re}(\psi))+Q(\mathrm{Im}(\psi))$, and $v_{kj}$ are elements of a (tranable) unitary $V=\sum_{kj}v_{kj}\ket{k}\bra{j}$, by using $VU_\psi$ as $U$ in Theorem \ref{NTCAalg}.
The amplitudes of this state can be regarded as the outputs of a single-layer neural network with input datum $\{\psi_k\}$.

While $U_\psi$ naturally includes the amplitude encoding unitary defined in Eq. (\ref{uni.data}),
it is notable that $U_\psi$ can be any encoding of a classical data $\Vec{x}$ to quantum state $\ket{\psi(\Vec{x})}$ such as the ones proposed in quantum machine learning literatures \cite{Mitarai_2018, Schuld_2019, Havl_ek_2019, farhi2018classification, glick2021covariant, Kusumoto2021}.
Moreover, we believe that a more important situation for NTCA is when the input state $\ket{\psi}$ has intrinsic quantumness.
For example, we might be able to learn properties of a certain family of Hamiltonians $H$ by directly feeding a time-evolved state $e^{iHt}\ket{0}$ or an approximate ground state generated via e.g. variational quantum eigensolver \cite{peruzzo2014variational} to a neural network.

The generated state has a variety of possible applications.
A straightforward application would be the classification.
To classify the data into $N_c$ classes, we measure certain $\lceil\log_2 N_c\rceil$ qubits of the output state (\ref{eq:quantumdata_nn}).
Note that in this case, we do not need to output a full vector like Eq. (\ref{eq:quantumdata_nn}) and can exploit the fact that we can output a state like $\sum_{k=1}^{N_c}F(\sum_{j=1}^N v_{kj}\psi_j)\ket{k}$.
Another possible application is to use the output state in Eq. (\ref{eq:quantumdata_nn}) for a generative model \cite{NIPS2001_7b7a53e2}.
The purpose of generative models is to create a machine that can sample from a probability distribution that is close to the distribution of data.
Since it has been shown that sampling from a certain family of quantum states is classically hard \cite{Lund2017}, it is a promising direction to explore the usefulness of quantum computers for this task. 
In contrast to the previous works \cite{PhysRevA.98.012324,romero2019variational, Zoufal_2019}, our algorithm can provide nonlinearity and it may have better performance compared with them.
Finally, we mention that the transformation can be considered as a quantum feature map \cite{Schuld_2019, Havl_ek_2019} that takes a quantum data $\ket{\psi}$ as an input and maps the state nonlinearly to create a quantum feature vector $\ket{\Psi(\psi)}=\sum_{k=1}^N F\left(\sum_{j=1}^Nv_{kj}\psi_j\right)\ket{k}$.
The inner product of the feature vectors, $\braket{\Psi(\psi_1)|\Psi(\psi_2)}$, can be exploited for quantum kernel methods \cite{Schuld_2019, Havl_ek_2019}, which constructs a machine learning model based on the inner products of data in a feature space.
We can also perform quantum circuit learning \cite{Mitarai_2018} by applying a trainable unitary to $\ket{\Psi(\psi)}$.

It is also possible to obtain the values of the $N_1(\leq N)$ dimensional nodes, i.e., $\left\{F\left(\sum_{j}v_{kj}\psi_j\right)\right\}_{k=1}^{N_1}$, by using amplitude estimation \cite{2000quant.ph..5055B}.
The amplitude estimation essentially allows us to estimate the absolute value $|a|$ of an amplitude $a$ in a quantum state $\ket{\phi} = a\ket{0}+\cdots$ with an additive error $\beta$ by $\mathcal{O}(1/\beta)$ uses of a quantum circuit that creates $\ket{\phi}$.
Note that $a$ itself can also be estimated if we can construct quantum circuits that create a state in the form of $\frac{1+a}{2}\ket{0}+\cdots$ and $\frac{1+ia}{2}\ket{0}+\cdots$.
In this case, we can separately estimate $|1+a|$, $|1+ia|$ and $|a|$ from which $\mathrm{Re}(a)$ and $\mathrm{Im}(a)$ can be reconstructed.
Since we can easily construct a quantum circuit that implements the functions $\frac{1+F}{2}$ and $\frac{1+iF}{2}$ instead of $F$ by QSVT, we can estimate each of $\left\{F\left(\sum_{j}v_{kj}\psi_j\right)\right\}_{k=1}^{N_1}$ with an additive error $\beta$ with $\mathcal{O}\left(\sqrt{N_1}/\beta\right)$ uses of the circuit in Fig. \ref{fig:3}. 
To estimate all of them, we need to repeat it for $N_1$ times and thus require $\mathcal{O}\left(N_1^{3/2}/\beta\right)$ uses of the circuit.

Finally, as a natural extension, we consider how to implement a multi-layer neural network. 
We first consider how to achieve the two-layer neural network.
More concretely, we wish to prepare a state in the form of,
\begin{align}\label{eq:quantumdata_twolayer_nn}
    \sum_{l=1}^{N_2} F\left[\sum_{k=1}^{N_1} v_{lk}^{(2)}F\left(\sum_{j=1}^N v_{kj}^{(1)}\psi_j\right)\right]\ket{l},
\end{align}
$V^{(\ell)}=\sum_{k,j} v_{kj}^{(\ell)}\ket{k}\bra{j}$ are tranable unitaries connecting the $(\ell-1)$-th and $\ell$-th layers, that is, we take input vector $\{\psi_j\}$ as zeroth layer.
We assumed the $\ell$-th layer has $N_\ell$ nodes.
Let $\mathcal{A}_1$ be the quantum circuit for a single-layer neural network, i.e., the circuit in Fig. \ref{fig:3} with $U=V^{(1)}U_\psi$.
It is easy to see that the state of Eq. (\ref{eq:quantumdata_twolayer_nn}) can essentially be constructed by using $\left(\mathcal{I}_{5}\otimes V^{(2)}\otimes \mathcal{I}_{n+1}\right)\mathcal{A}_1$ as the state preparation oracle in Theorem \ref{NTCAalg}.
Detailed analysis of this process is presented in Appendix \ref{app.nn.multi}.
$\ell$-layer neural network can be achieved by repeating the above steps $\ell-1$ times.
As shown in Appendix \ref{app.nn.multi}, a quantum state which corresponds to an $\ell$-layer neural network can be generated by  $\mathcal{\tilde{O}}\left(d^\ell\sqrt{N_1N_2\cdots N_{\ell-1}}\right).$

\section{Conclusion}

We have deﬁned a task called NTCA considering how to apply nonlinear functions onto the amplitudes of a quantum state. 
To this end, 
we introduced \textit{block-encoding of amplitudes},
which encodes amplitudes of a quantum state on a block element of 
a unitary operator by using the state preparation oracle.
This allows us to treat complex amplitudes of a quantum state 
as singular values of a matrix encoded into a unitary operator and hence to apply QSVT 
for NTCA.
Since the proposed mapping from a state preparation oracle 
to a block-encoding unitary is very general,
we expect that it can be applied widely when we want to process 
a quantum state in such a way that cannot be achieved by a simple unitary transformation.
As a concrete example of such an application, 
we proposed to use our algorithm to implement neural networks 
on quantum computers, where complex amplitudes 
are used as nodes for the neural networks and 
nonlinear transformations, which cannot be realized by a unitary operation,
are performed by the proposed NTCA.
From scaling analysis, 
our result implies that there could be 
a complexity-theoretic benefit to use 
a quantum computer to process and perform 
a nonlinear transformation on a high dimensional input data,
while it still relies on several assumptions such as
the state preparation oracle.
It is an interesting open question whether the proposed nonlinear transformation or quantum neural network can provide an advantage in classical machine learning tasks.
We think that it is, at least, inevitable to apply quantum machine learning for quantum data
since a systematic nonlinear transformation is essentially missing in quantum systems
without the proposed method.

\paragraph*{Note-added.}
The first preprint version of this manuscript was released in July 2021. Since then, there have been several further developments.
A similar task was considered in the parameterized quantum circuit setting \cite{PhysRevResearch.5.013105}.
By using the importance-weighted method and for functions like $\tanh(x)$, one can remove the dependency on the input dimension $N$ \cite{rattew2023nonlinear}. 
Applications like quantum machine learning, state preparation, and maximum finding have been investigated \cite{GonzalezConde2024efficientquantum, guo2024quantum, rattew2023nonlinear, roga2023fully}.

\begin{acknowledgments}
NG is supported by Hirose Foundation.
KM is supported by JST PRESTO Grant No. JPMJPR2019 and JSPS KAKENHI Grant No. 20K22330.
KF is supported by JSPS KAKENHI Grant No. 16H02211,  JST ERATO JPMJER1601, and JST CREST JPMJCR1673.
This work is supported by MEXT Quantum Leap Flagship Program (MEXT QLEAP) Grant Number JPMXS0118067394 and JPMXS0120319794.
We also acknowledge support from JST COI-NEXT program.
\end{acknowledgments}

\appendix

\section{Details for Theorem \ref{BCA}\label{appendix:a}}

\subsection{Derivation of Eq. (\ref{eq:op_W})}\label{app.op.W}
\begin{enumerate}
    \item Apply the oracle $U$ on data qubits, the state will be      
        \begin{align}
        \ket{k}_{\mathrm{ad}}\left(\sum_{j=1}^Nc_j\ket{0}^{\otimes a}\ket{j}\right)_{\mathrm{da}}\ket{0}_B.
        \end{align} 
    \item Perform Hadamard gate on the $B$ ancilla qubit and controlled-$U$ gate with $B$ as controlled qubits, we prepare the state
    \begin{align}
        \ket{k}_{\mathrm{ad}}
        \biggl(&\bigl(\sum_{j=1}^Nc_j\ket{j}\bigl)_{\mathrm{da}}\ket{0}_B+\ket{0}^{\otimes n}_{\mathrm{da}}\ket{1}_B\biggl).
    \end{align}
    \item
    Apply Toffoli gates with address qubits and $B$ ancilla qubit as controlled gates and Hadamard gate on the $B$ ancilla qubits, we have 
    \begin{align}\label{eq.psi.k}
        \ket{k}_{\mathrm{ad}}\ket{\Psi_k}_{\mathrm{da,B}}
        :=&\frac{\ket{k}}{2}_{\mathrm{ad}}\biggl(\bigl(\sum_{j=1}^N c_j\ket{j}+\ket{k}\bigl)_{\mathrm{da}}\ket{0}_{\mathrm{B}}\notag\\
        &+\bigl(\sum_{j=1}^Nc_j\ket{j}-\ket{k}\bigl)_{\mathrm{da}}\ket{1}_{\mathrm{B}}\biggl).
    \end{align}
\end{enumerate}

\subsection{Eigenvalues and eigenvectors of $G$ \label{app.op.G}}

Here, we derive the eigenvalues and eigenvectors of $G$ defined in Eq. \ref{eq:def_G}.
It can be seen that $G$ has invariant two-dimensional subspaces spanned by $\{\ket{k}\ket{\Psi_{k0}},\ket{k}\ket{\Psi_{k1}}\}_{k=1}^N$ for each $k\in\{1,...,N\}$.
Let $\overline{Z_B},\overline{WS_0W^\dagger},\overline{G}$ be the operations $Z_B,WS_0W^\dagger$ and $G$ acting only on these subspaces.

The matrix representation of $\overline{Z_B},\overline{WS_0W^\dagger}$ can be written as,
\begin{align}
    \overline{Z_B}=&|k\rangle|\Psi_{k0}\rangle\langle\Psi_{k0}|\langle k|-|k\rangle|\Psi_{k1}\rangle\langle\Psi_{k1}|\langle k|\\
    \overline{WS_0W^\dagger}=&(1-2\alpha^2_k)|k\rangle|\Psi_{k0}\rangle\langle\Psi_{k0}|\langle k|\notag\\
    &+(1-2\beta^2_k)|k\rangle|\Psi_{k1}\rangle\langle\Psi_{k1}|\langle k| \notag\\
    &-2\alpha_k\beta_k \left(|k\rangle|\Psi_{k1}\rangle\langle\Psi_{k0}|\langle k|+|k\rangle|\Psi_{k0}\rangle\langle\Psi_{k1}|\langle k|\right).
\end{align}
Thus, we have, 
\begin{align}
    \overline{G}=&\overline{WS_0W^\dagger}\overline{Z_B}\\
    =&(1-2\alpha^2_k)|k\rangle|\Psi_{k0}\rangle\langle\Psi_{k0}|\langle k|\notag\\
    &-(1-2\beta^2_k)|k\rangle|\Psi_{k1}\rangle\langle\Psi_{k1}|\langle k| \notag\\
    & -2\alpha_k\beta_k (|k\rangle|\Psi_{k1}\rangle\langle\Psi_{k0}|\langle k|-|k\rangle|\Psi_{k0}\rangle\langle\Psi_{k1}|\langle k|).
\end{align}
Therefore, eigenvectors of $\overline{G}$ can be written as 
\begin{align}\label{eq:eigenvector_G_overline}
    \ket{k}_{\mathrm{ad}}\ket{\Psi_{k\pm}}_{\mathrm{da,B}}=\frac{1}{\sqrt{2}}\ket{k}_{\mathrm{ad}}(\ket{\Psi_{k0}}\pm i\ket{\Psi_{k1}})_{\mathrm{da,B}},
\end{align}
with the corresponding eigenvalue $-x_k\pm i\sqrt{1-x_k^2}$.
$G$ has the same eigenvalues and eigenvectors.

\subsection{One- and two-qubit gate counts \label{app.num.gates}}

To construct $\tilde{U}$ as in Theorem \ref{BCA}, we use controlled-$G$ and controlled-$G^\dagger$ one time, which consists of controlled-$H$, controlled-$Z$, Toffoli gates on four qubits, and controlled-$S_0$ gates.
Decomposition of Toffoli gates is described in \cite{nielsen_chuang_2010}.
Notice that here we use Toffoli gates $\mathcal{O}(n)$ times.
The (controlled) conditional phase shift gate $S_0=\mathcal{I}_{n+2}-2|0\rangle\langle 0|$ on $n$ qubits can be performed with a Toffoli gate on $(n+2)$ qubits and two Hadamard gates.
As described in \cite{cite-key}, Toffoli gate on $(n+2)$ qubits can be performed with $\mathcal{O}(n)$ gates with an additional ancilla qubit.
Therefore, in total, we use controlled-$U$ and controlled-$U^\dagger$ four times and $\mathcal{O}(n)$ one- and two-qubit gates for Theorem \ref{BCA}.

\section{Detailed error analysis for Theorem \ref{NTCAalg} \label{app.err.any}}

Note that  $\mathcal{P}$ is a $(1,4,\epsilon/(16\gamma))$-block-encoding of a matrix whose eigenvalues are $P'(x_k)/(4\gamma)$.
For all $x_k\in [-1,1]$, we have 
\begin{align}
    \left|P'(x_k)/(4\gamma)-P''(x_k)\right|\leq \varepsilon/(16\gamma N),
\end{align}
where $P''(x_k)$ denotes the exact value achieved by Lemma \ref{QSVT}. 
By assumption, we have $|P(x_k)-P'(x_k)|\leq \varepsilon/(4N)$.
Therefore,
\begin{align}
    &|P(x_k)-4\gamma P''(x_k)| \notag\\
    \leq& |P(x_k)-P'(x_k)|+|P'(x_k)-4\gamma P''(x_k)|\notag\\
    \leq& |P(x_k)-P'(x_k)|+4\gamma|P'(x_k)/(4\gamma)- P''(x_k)|\leq \frac{\varepsilon}{2N},
\end{align}
where the inequalities comes from triangle inequality.
The same discussion applies for the imaginary part, i.e., $|Q(y_k)-4\gamma Q''(y_k)|\leq \varepsilon/(2N)$ for all $y_k\in [-1,1]$.
Finally, for all $x_k,y_k\in [-1,1]$ we have
\begin{align}
    |4\gamma P''(x_k)+4\gamma Q''(y_k) -P(x_k)-Q(y_k)|\leq \frac{\varepsilon}{N},
\end{align}
satisfying the error bound condition for NTCA.

\section{Polynomial approximation of tanh}\label{app.appro.tanh}
Here, we show that $P(x)=\tanh(x)$ can be approximated to an precision $\varepsilon$ with a degree $d=\mathcal{O}(\log(1/\varepsilon))$ polynomial $P'_d(x)$ in the sense that $|P'_d(x)-\tanh(x)|\leq\varepsilon$.
It suffices to use the simple Taylor series.
By Taylor expansion, 
\begin{align}
    \tanh(x)=\sum_{n=1}^\infty (-1)^{n-1}\frac{2^{2n}(2^{2n}-1)B_{2n}}{(2n)!}x^{2n-1},
\end{align}
where $B_{2n}$ is the Bernoulli number.
We take,
\begin{align}
    P'_d(x) = \sum_{n=1}^{d}(-1)^{n-1}\frac{2^{2n}(2^{2n}-1)B_{2n}}{(2n)!}x^{2n-1}.
\end{align}
Then, the error can be written as,
\begin{align}\label{eq.tanh.appro}
    |P'_d(x)-\tanh(x)| = \sum_{n=d+1}^\infty (-1)^{n-1}\frac{2^{2n}(2^{2n}-1)B_{2n}}{(2n)!}x^{2n-1}.
\end{align}
As mentioned in Ref. \cite{LEEMING1989124}, we have  
\begin{align}
    |B_{2n}|<  5\sqrt{\pi n}\left(\frac{n}{\pi e}\right)^{2n}.
\end{align}
Also, since $\left(n/e\right)^n\leq n!$, we have
\begin{align}
    \frac{2^{2n}(2^{2n}-1)B_{2n}}{(2n)!} &< 5\sqrt{\pi n}\left(\frac{4\frac{n}{\pi e}}{\frac{2n}{e}}\right)^{2n}\notag\\
    &=5\sqrt{\pi n}\left(\frac{2}{\pi}\right)^{2n}.
\end{align}
Further, since $\sqrt{d+1}\leq (\pi/2)^{d+1}$ for $d\geq 0$, we have
\begin{align}
    5\sqrt{\pi (d+1)}\left(\frac{2}{\pi}\right)^{2d+2}\leq 5\sqrt{\pi} \left(\frac{2}{\pi}\right)^{d+1}.
\end{align}
Therefore, the error in Eq. (\ref{eq.tanh.appro}) can be bounded by
\begin{align}
    &\sum_{n=d+1}^\infty (-1)^{n-1}\frac{2^{2n}(2^{2n}-1)B_{2n}}{(2n)!}x^{2n-1}\notag\\
    <& \sum_{n=d+1}^\infty \frac{2^{2n}(2^{2n}-1)B_{2n}}{(2n)!}x^{2n-1}\notag\\
    \leq& \sum_{n=d+1}^\infty 5\sqrt{\pi} \left(\frac{2}{\pi}\right)^{d+1}\notag\\
    =&\frac{5\sqrt{\pi}}{1-(2/\pi)}\left(\frac{2}{\pi}\right)^{d+1}.
\end{align}
It follows that the error bound of $\varepsilon$ can be achieved by setting $d= \mathcal{O}\left(\log\left(1/\varepsilon\right)\right)$.

\section{Generalization to multi-layer neural network\label{app.nn.multi}}
We first consider how to perform a two-layer quantum neural network.
For simplicity, we assume $|P(x)|,|Q(x)|\leq 1$ for all $x\in [-1,1]$ so that $\gamma=1$.
We also assume $P(x)$ and $Q(x)$ can exactly be expressed by a $d$-degree polynomial to avoid a complicated discussion about errors.
Let $\mathcal{A}_1$ be the quantum circuit that outputs the state
\begin{align}\label{eq.nn.two}
    \frac{1}{8\sqrt{N_1}}\sum_{k=1}^{N_1}F\left(\sum_{j=1}^N v_{kj}^{(1)}\psi_j\right)\ket{0}^{\otimes 5}\ket{k}_{\mathrm{ad}}\ket{0}^{\otimes n+1}_{\mathrm{da,B}}+\cdots.
\end{align}
Also, let $\tilde{\mathcal{A}}_1:=\left(\mathcal{I}_{5}\otimes V^{(2)}\otimes \mathcal{I}_{n+1}\right)\mathcal{A}_1$. Then,
\begin{align}
    \begin{split}
        &\tilde{\mathcal{A}}_1\ket{0}^{\otimes 2n+6}=\\
    &\frac{1}{8\sqrt{N_1}}
    \sum_{l=1}^{N_2} \left[\sum_{k=1}^{N_1} v_{lk}^{(2)}F\left(\sum_{j=1}^N v_{kj}^{(1)}\psi_j\right)\right]\ket{0}^{\otimes 5}\ket{l}_{\mathrm{ad}}\ket{0}^{\otimes n+1}_{\mathrm{da,B}}+\cdots.
    \end{split}
\end{align}
$\tilde{\mathcal{A}}_1$ can be considered as the state preparation oracle for Theorem \ref{NTCAalg} to achieve the two-layer neural network.
However, before doing so, we remove the coefficient $1/(8\sqrt{N_1})$ by uniform singular value amplification \cite{10.1145/3313276.3316366, low2017hamiltonian}.
To do so, first note that, by Theorem \ref{BCA}, we can construct a block-encoding $\tilde{G}^{(2)}$ of a matrix $-\frac{1}{2}\left(G^{(2)}+(G^{(2)})^\dagger\right)$ whose eigenvalues are \begin{align}
    \left\{\frac{1}{8\sqrt{N_1}}\sum_{k=1}^{N_1} v_{lk}^{(2)}F\left(\sum_{j=1}^N v_{kj}^{(1)}\psi_j\right)\right\}_{k=1}^{N_1}.
\end{align} 
The coefficient $1/(8\sqrt{N_1})$ from these eigenvalues can be removed by using uniform singular value amplification which uses $\tilde{A}_1$ and $\tilde{A}^\dagger_1$ $\tilde{\mathcal{O}}\left(\sqrt{N_1}\right)$ times.
Let $\tilde{\mathcal{A}}_1'$ be the circuit constructed by the above procedure, i.e.,
\begin{align}
    \begin{split}
        &\tilde{\mathcal{A}}_1'\ket{0}^{\otimes 2n+6} \\
    &=\sum_{l=1}^{N_2} \left[\sum_{k=1}^{N_1} v_{lk}^{(2)}F\left(\sum_{j=1}^N v_{kj}^{(1)}\psi_j\right)\right]\ket{0}^{\otimes 5}\ket{l}_{\mathrm{ad}}\ket{0}^{\otimes n+1}_{\mathrm{da,B}}+\cdots.
    \end{split}
\end{align}
After this step, we use Theorem \ref{NTCAalg} on $\tilde{\mathcal{A}}_1'$ to achieve the state 
\begin{align}
    \frac{1}{8\sqrt{N_2}}\sum_{l=1}^{N_2} F\left[\sum_{k=1}^{N_1} v_{lk}^{(2)}F\left(\sum_{j=1}^N v_{kj}^{(1)}\psi_j\right)\right]\ket{0}^{\otimes 5}\ket{l}_{\mathrm{ad}}\ket{0}^{\otimes n+1}_{\mathrm{da,B}}+\cdots.
\end{align}
Counting the $U_\psi$ operations used in this whole procedure, this state can be prepared with $\mathcal{\tilde{O}}\left(d^2\sqrt{N_1}\right)$ uses of $U_\psi$.

An $\ell$-layer neural network can be achieved by repeating the above procedure $\ell-1$ times, and it uses
$\mathcal{\tilde{O}}\left(d^\ell\sqrt{N_1N_2\cdots N_{\ell-1}}\right)$ times of $U_\psi$.
If we want to further obtain the values of each amplitude with an additive error $\beta$, we perform the amplitude estimation on the resulting state. This would require repeating the whole circuit for $\mathcal{O}\left(N_\ell^{3/2}/\beta\right)$ times, and the total number of $U_\psi$ is calculated to be
$\mathcal{\tilde{O}}\left(d^\ell N_\ell\sqrt{N_1N_2\cdots N_{\ell}}/\beta\right)$.

\bibliography{apssamp}

\end{document}